# Topology, Geometry, and Stability: Protein Folding and Evolution

Walter Simmons
Department of Physics and Astronomy
University of Hawaii at Manoa

Joel L. Weiner
Department of Mathematics
University of Hawaii at Manoa

ABSTRACT

The protein folding problem must ultimately be solved on all length scales from the atomic up through a hierarchy of complicated structures. By analyzing the stability of the folding process using physics and mathematics, this paper shows that features without length scales, i.e. topological features, are potentially of central importance.

Topology is a natural mathematical tool for the study of shape and we avail ourselves of that tool to examine the relationship between the amino acid sequence and the shapes of protein molecules. We apply what we learn to conjectures about their biological evolution.

Introduction

Some features of protein folding are topological, others are geometrical.  The mathematical distinction between topology and geometry is clear and is important for understanding both folding and protein evolution.  Geometric features are those that depend upon a scale length; such as anything that involves forces, energy, bond angles, and most noise. Topological features do not depend upon scale lengths.  The central importance of this distinction arises in the stability of processes; topological features may be less sensitive to perturbations (which are usually geometrical).  We suppose that during billions of years of evolution, natural selection has preserved some biological processes based upon the reliability and noise-resistance required for successful reproduction in an ever-changing environment.  In this paper we explore the possible role of topology in protein folding and evolution.  Our analysis of protein folding from a topological viewpoint arrived at the following qualitative and semi-quantitative results: that topological constraints can impact the secondary structure down to the single residue level, that possible topological effects can increase the dispersal of the code (sequence to shape) over the molecule, that there is a natural explanation of dynamical Levinthal pathways (1).  Of course, the Levinthal pathways go to explain the speed and energy-efficiency of folding.  It is suggested that similar analysis applied to protein evolution has implications for protein family structures.

A traditional approach to protein folding theory is directly or indirectly equivalent to the following: generalized coordinates are devised, and a Lagrangian, $L$ is developed. The action follows as $S \equiv \int_0^t L dt$ and the stability of the action under variation of paths leads to the Euler-Lagrange differential equations. The differential equations are then simulated.

The relative roles of topology VS geometry have drawn considerable interest in physics in the past few decades, however the activity has been in sub-fields quite different from molecular biology and the methods used generally do not transfer over. Here we treat folding as a mechanical wave propagation phenomenon and proceed in analogy with classical optics, which follows a strong parallel with mechanics as first discovered by Hamilton (2).

To explore this topology/geometry distinction in protein physics, we need a theory of stability that can be applied to the folding process and to the native structure. Thermodynamics is obviously important but these folding processes are generally non-equilibrium. Catastrophe theory (3),(4),(5) is a topological theory of stability of great generality which beautifully augments mechanics, quantum mechanics, and thermodynamics.

One of the issues we address is this: how much detail can be controlled by topological factors alone? As we shall show, below, the topology of a protein molecule (i.e. aspects independent of length scale) can strongly influence shape of the secondary structure even at the individual residue scale.

Stability in Catastrophe Theory

Catastrophe theory, the name notwithstanding, is a mathematical theory of stability.  In mathematics, the theory describes a family of functions which have the same shape.  The family is described in terms of a generating function which depends upon state-variables and control parameters.  In physics applications, the state-variables describe the behavior of a physical system in the same way that generalized coordinates are used in mechanics.  The control parameters usually describe the measurements on the physical system.  When the system behavior is measured using a sub-set of the control parameters, the projection onto the limited space of control variables is not influenced by small perturbations of the generating function.

We note that each of the generating functions that we consider has a single normal form, which is polynomial in the state-variables and linear in the control parameters († †).

Several excellent and complete books on catastrophe theory are available (3),(4).  No short introduction to catastrophe theory as a theory of stability in physics seems to be available so this section serves that purpose.

To proceed, we need to name the family of functions.  Typically, the family is portrayed in the literature as a generating function.  To illustrate, we shall limit ourselves, for the moment, to generating functions of a single state variable, $X$.

The simplest generating function in this case is the fold,

$$A_2 = \frac{1}{3}X^3 + a_1 X$$

The generating function for the cusp $A_{\pm 3}$, with state-variable $X$ and control parameters $a_1, a_2$ is

$$A_{\pm 3} = \pm \frac{1}{4} X^4 + \frac{1}{2} a_2 X^2 + a_1 X$$

(Note: some authors prefer to call the generating functions 'catastrophes').

Contrary to what is sometimes written, there is an infinite tower of these single state-variable generating functions with the highest power of the state variable being $X^{k+2}$ and the dimension of the control space $k$ ; that is a set of $k$ control parameters, $a_k, a_{k-1}, ..., a_1$.

Second order critical points are places where the first and second derivatives of the generating functions vanish. In the above example, there is a triply degenerate critical point at $(X; a_1, a_2) = (0; 0, 0)$

$$\frac{\partial A_{\pm 3}}{\partial X} = X^3 + a_2 X + a_1 = 0$$

$$\frac{\partial^2 A_{\pm 3}}{\partial X^2} = 3X^2 + a_2 = 0$$

$$\frac{\partial^3 A_{\pm 3}}{\partial X^3} = 6X = 0$$

When this occurs in geometrical optics, the diffraction pattern on a screen is given by,

$$\left(\frac{a_2}{3}\right)^3 + \left(\frac{a_1}{2}\right)^2 = 0$$

All the light passes through this cusp shaped curve, and this cusp shape is stable against perturbations of the generating function.

At the end of this section, we summarize a beautiful application of catastrophe theory to optics which, it is hoped, will make some of these generalities clearer. In that example, catastrophe theory applies to the optical path length, the state-variables parameterize the ray paths through the lens, and the control parameters refer to the position of a screen on which the diffraction pattern is viewed. Stability refers to the fact that small changes in the lens or the position of the screen do not change the shape of the diffraction pattern on the screen.

In addition to stability, two important results that emerge from the application of catastrophe theory to mechanical systems are that all of the energy propagating in a system with a catastrophe passes through the catastrophe and that there are a very few state-variables needed to describe the physics and most other variables can be disregarded. There are many other results to be obtained from catastrophe theory, many of which go by the name 'catastrophe flags' (3).

A general pattern of catastrophe theory applications in physics starts by identifying two conditions. The first condition is that there is a physical system (usually multi-component) that is stable. The second condition is that a critical point of at least second order appears in the dynamics. In potential theory the first condition is often met by thermodynamic equilibrium (6),(7),(8). The presence of a second order phase transition provides the second condition. In optics and in mechanics the stability can arise from the fact that the action is stable against perturbations of the path of motion. The second condition can arise if there is a caustic or kinetic recurrence in the motion (5),(2),(9).

We turn, finally, to a summary of the topology.

We have already mentioned that the stability has a topological origin. The dimensions of spaces, manifolds, etc. form essential quantities for topology.

Let $k$ stand for the dimension of the space of control parameters. E.g., $k=1$ for the fold and $k=2$ for the cusp $A_{\pm 3}$.

Where we limit ourselves to the case of a single state-variable and consider only the elementary catastrophes of Thom, $n$ would be the largest exponent of a state-variable. That would be $n=4$ for the cusp.

In the general case, there are multiple independent state-variables and a sub-set of them will have second order critical points. Let the number of such variables be $\ell$.

For topological stability, which is stability of shape as described, the following minimum value of $k$ is required (3),

$$k \geq \frac{\ell(\ell+1)}{2}$$

All of the single state-variable generating functions have $\ell=1$ and the minimum dimension of the control space is $k \geq 1$. However, for larger values of $k$ higher powers of $x$ occur, that is $x^{k+2}$.
As an example of all this, we shall summarize an optics experiment on a caustic described, (with accompanying theoretical analysis), in a paper by Berry, Nye, and Wright (10). One reason for selecting this example is that the most important results were obtained from optics without specific reference to catastrophe theory; the catastrophe theory, which is very general, was presented in parallel with the optics. Another motive is that in optics, the path-length of light rays to the diffraction pattern is always stable against small perturbations, and that it is stable to second order up to meeting a caustic.

We list specific features of this experiment and explain how each is related to catastrophe theory, which provides a much more general picture than optics alone.

Berry, Nye, and Wright engineered a triangular lens which produces an optical caustic. The caustic has a framework or skeleton that is in the shape of Thom's elliptic umbilic catastrophe plus a diffraction pattern decorating the framework. (They refer to the caustic as a 'diffraction catastrophe').
From standard optics, the calculated far-field pattern is

$$E(x,y,z) = \frac{1}{2\pi} \int d\xi \int d\eta \exp iP(x,y,z,\eta,\xi)$$

$$= \frac{1}{2\pi} \int d\xi \int d\eta \exp i[\xi^3 - 3\xi\eta^2 - z(\xi^2 + \eta^2) - x\xi - y\eta],$$

where $P(x,y,z,\eta,\xi)$ is the optical path length and the integrations were taken over the lens. The variables $\xi$ and $\eta$ are known as state-variables and the coordinate frame $(x,y,z)$ is known as the control space.

The critical points are,

$$\frac{\partial P}{\partial \xi} = 3(\xi^2 - \eta^2) - 2z\xi - x = 0$$

$$\frac{\partial P}{\partial \eta} = -6\xi\eta - 2z\eta - y = 0$$

The caustic is defined as the singularities (degenerate critical points) found by the vanishing of the Hessian determinate,

$$(\frac{\partial^2 P}{\partial \xi^2})(\frac{\partial^2 P}{\partial \eta^2}) - (\frac{\partial^2 P}{\partial \xi \partial \eta})^2 = 0$$

This catastrophe, known as $D_{-4}$, has been explained in detail in (3). With all of the control parameters set to zero, the generating function has a four-fold degenerate critical point at the $(\eta,\xi)$ origin.

The overall picture is of a family of optical rays that have critical points of various orders in the control space.

Kirchhoff-Fresnel diffraction theory tells us that each point in the far-field receives light from every point in the object field but catastrophe theory makes a stronger statement. If a catastrophe caustic is present, then <u>all</u> of the light falls upon the catastrophe. Typically in catastrophe theory only a few state-variables are sufficient to describe the physics, while many other variables are not important; this feature of the theory involves both topology and geometry. The distinction is that state-variables that relate to a second order critical point dominate the physical behavior.

The most important topological aspect of catastrophe theory arises in the stability against perturbations.  Full expositions of this subject, which are available in the standard references, are rather involved and will not be repeated here.  However, we can state the underlying topological result.

A key point is that the shape of the diffraction pattern, ($E(x, y, z)$, above), is stable against perturbations of the generating function.

As is usually the case, the control space describes the way the phenomenon is known to experimenters or observers.  Here the coordinates $(x, y, z)$ represent the position of point on a projection screen.

The generating function (action) $P(x, y, z, \eta, \xi)$ has $\ell = 2$, and the dimension of the control space is $k = 3$ for the $(x, y, z)$ space.  Note, from topological considerations, it is necessary that $k \geq 3$ in order to have stability.

As a technical note, we mention the concept of transversality (3),(4).  If two manifolds, P and Q, intersect in a space of dimension p, the intersection is said to be transversal if it is topologically invariant under small perturbations. The exact statement refers to the tangent spaces of the manifolds at the intersection and is: at any point of the intersection, the space of vectors spanned by the tangent spaces of the manifolds at that point have dimension $\geq p$.

In the optics experiment, stability of the projection follows from the fact that the Hessian of the generating function of the path length, as a function of control parameters, is transversal to a manifold of symmetric matricies.

The hierarchical geometry and topology of the catastrophes is of interest. When a catastrophe such as the elliptic-umbilic, above, is examined in detail, folded sheets and cusped sheets are seen. That is, lower level catastrophes are visible in the higher level forms. For example, in the infinite tower of catastrophes with a single state-variable, displaying the critical points reveals equations for the fold and the cusp. High power catastrophes in the tower contain all of the lower power catastrophes.

To sum up this section, the analysis of the optics experiment and underlying theory illustrated that a number of physical properties typical of the very general catastrophe theory can be obtained directly from the physics in a specific case. It showed that the diffraction catastrophe entails two state-variables and three control parameters, it is stable, and all the light in the far field falls on the catastrophe.

Protein Folding and Evolution

We view the folding process as occurring through an ensemble of mechanical waves (mostly torsion waves), moving along the molecule, and starting from the denatured molecule in solution and arriving at a state where the secondary and tertiary structures are mostly defined. What follows is an annealing process at atomic-dimensions.

Our fundamental conjecture (9) is that, by selecting for protein molecules that fold stably in the presence of perturbations, natural selection has picked out internal parameters that lead to the presence of a catastrophe (or catastrophes) in the action of the folding process. All of the free energy follows a pathway to the projection of the catastrophe upon the space of dihedral angles.

Several semi-quantitative physical consequences follow immediately: Levinthal pathways (1) can be defined in terms of state-variables and control parameters in the action, the speed and uniqueness of folding follows from the efficient free energy utilization along the folding path to the projection of the catastrophe, and, most importantly, the stability arises via catastrophe theory, as in the optics experiment (9).

Next we examine additional aspects of folding that follow from our topological analysis.

The action is a scalar time-integral over the motion and includes contributions from all structures in the molecule.

We know the shape of the molecule by its position in the space of dihedral angles. We provisionally adopt the assumption that the $k$ control parameters are dihedral angles. We don't know how many state-variables there may be. With just one state-variable, the topological theory outlined above leads to the requirement for just one control parameter (fold catastrophe). This minimal version of our theory indicates that topological considerations may impact the folding at the single residue level.

Note, however, the $k$ value is a minimum; more control parameters can be involved in that the Hessian of the generating function with respect to state-variables will depend upon these additional parameters.

If there are more state-variables, or if there are high powers of terms in the generating function, then the number of control parameters must expand; if the Hessian has nullity $\ell$, (nullity is the number of zero eignevalues) then the smallest number of control parameters would increase quadratically with $\ell$.

In either case (more control parameters or more state-variables), many dihedral would code for a secondary or tertiary structure; the code (sequence to shape) would be dispersed over many residues (not necessarily consecutive).

As an environment for wave propagation, the molecules are complicated and dynamic on many length scales. As mentioned, it is known from the field of wave propagation in random media that if the scattering structures are smooth then catastrophes emerge quickly. If the random structures have sharp edges or points, then diffeomorphisms of the generating function do not apply and premises of catastrophe theory do not apply.

(As a technical note, high dimension catastrophes have additional parameters called 'moduli' (11), which we do not consider here.)

We next assume that reliable and stable protein folding is selected during mutation and consequent protein evolution.

Mutations can impact the shape of the molecule in multiple ways. A catastrophe may move to a new position in the space of dihedral angles and, perhaps, the final annealing may change significantly, there can be changes in the importance of various control parameters, and new state variables may appear.

The introduction of new control parameters can disperse the code over a larger number of residues.

The introduction of a higher power in the generating function may result in appearance of a new hierarchical level of structure that subsumes ancestral structures. (This has been observed in catastrophe optics (12) and may be a hierarchic pattern in protein folding (13)).

These evolutionary mechanisms are compatible with a common assumption that the steps are represented by multiplicative probabilities and therefore log-normal statistical distributions of protein families are expected.

(††) From the mathematical perspective we do not want to distinguish between generating functions if one can be obtained from the other by a change of independent or dependent coordinates; such generating functions are regarded as equivalent. Among all equivalent generating functions there may be some which have a particularly simple form; for the generating functions we consider there is, in general, a single normal form. When we consider generating functions in this paper we deal with their normal forms.

References

1. Levinthal, C. *Extrait du Journal de Chimie Physique*, 1968, page 44

2. Gray,C.G &. Taylor, E. When Action is Not Least, Am. J. Phys. 75, 434 (2007);

3. Gilmore, R. Catastrophe Theory for Scientists and Engineers. John Wiley & Sons, 1981.

4. Poston,T. & Stewart I., Catastrophe Theory and its Applications, Pitman Publishing Limited, 1978

5. Berry, M.V, & Upstill,C., Catastrophe Optics, Progress in Optics, 1980, Vol. XVIII, 257-346.

6. Kusmatsev, F.V. Physics Reports 188, 1-36 (1994).

7. Wales, D. & Head-Gordon, T., Evolution of the Potential Energy Landscape with Static Pulling Force for Two Model Proteins, J. Phys. Chem. B (2012), 116, 8394−8411

8. Schulman, L.S. & Revzen,M., Phase Transitions as Catastrophes, Collective Phenomena 1, 43-49, (1972)

9. Simmons,W. & Weiner, J.L., The Principle of Stationary Action in Biophysics: Stability in Protein Folding, Journal of Biophysics,

Volume 2013 (2013), Article ID 697529

10. Berry, M.V, Nye,J.F., Writht,F.J., The Elliptic Umbilic Diffraction Catastrophe, Philosophical Transactions of the Royal Society, (1979), Vol. 291, pp 32-36 .

11. Schulman, L.S., "Phase Transitions with Several Order Parameters", Physica D, 89A, 597-604 (1977).

12. Nye, J.F., The catastrophe optics of liquid drop lenses, Proc. R. Soc. Lond. <u>A 403</u>, 1-26 (1986).

13. Baldwin, R.L. & Rose, G.D., "Is protein folding hierarchic? I. Local structure and peptide folding", Trends Biochem Sci, <u>24</u>, 26-35 (1999).